\documentclass{article}

\usepackage{arxiv}

\usepackage[utf8]{inputenc} 
\usepackage[T1]{fontenc}    
\usepackage{hyperref}       
\usepackage{url}            
\usepackage{booktabs}       
\usepackage{amsfonts}       
\usepackage{nicefrac}       
\usepackage{microtype}      
\usepackage{lipsum}		
\usepackage{graphicx}
\usepackage{doi}
\usepackage{biblatex}
\usepackage{amsmath}
\usepackage{caption}
\usepackage{subcaption}
\usepackage[export]{adjustbox}
\usepackage{wrapfig}

\title{A Novel Approach to Encode Two-Way Epistatic Interactions Between Single Nucleotide Polymorphisms}

\author{Nathaniel Gunter \\
	Department of Radiology\\
	Mayo Clinic\\
	Rochester, MN\\
	\texttt{gunter.nathaniel@mayo.edu} \\
	\And
    Prashanthi Vemuri \\
	 Department of Radiology \\
    	Mayo Clinic \\
     Rochester, MN\\
 	\AND
	 Vijay Ramanan \\
	 Department of Neurology \\
        Mayo Clinic \\
	 Rochester MN \\
	 \And
        Robel K. Gebre \\
	Department of Radiology\\
	Mayo Clinic\\
	Rochester, MN \\	 
}

\date{15 June, 2023}

\renewcommand{\shorttitle}{Encoding Two-Way Epistatic Interactions}

\hypersetup{
pdftitle={A Novel Approach to Encode Two-Way Epistatic Interactions Between Single Nucleotide Polymorphisms},
pdfsubject={q-bio.NC, q-bio.QM},
pdfauthor={Nathaniel Gunter},
pdfkeywords={genetics, epistatic interactions},
}

\begin{document}
\maketitle

\begin{abstract}
\textbf{Introduction}

Modelling gene-gene epistatic interactions when computing genetic risk scores is not a well-explored subfield of genetics and could have potential to improve risk stratification in practice. Though applications of machine learning (ML) show promise as an avenue of improvement for current genetic risk assesments, they frequently suffer from the problem of two many features and to little data. We propose a method that when combined with ML allows information from individual genetic contributors to be preserved while incorporating information on their interactions in a single feature. This allows second-order analysis, while simultaneously increasing the number of input features to ML models as little as possible.
\textbf{Methods}

Data was retrieved from the Aging and Dementia Neuroimaging Initiative, consisting of genotype, age, sex, genetic principal components, and amyloid centiloid metrics. The data were then put through one of three genetic interaction methods, and the resultant interaction features were used as inputs for a series of machine learning models. The $r^2$ association between actual and predicted amyloid centiloid was collected for each of a hundred runs and used to construct distributions for each set of generated features.

\textbf{Results}

We presented three methods that can be utilized to account for genetic interactions. We found that interaction methods that preserved information from the constituent SNPs performed significantly better than the simplest interaction method. Since the currently available ML methods are able to account for complex interactions, utilizing raw SNP genotypes alone is sufficient because the simplest model outperforms all the interaction methods

\textbf{Discussion}

Given that understanding and accounting for epistatic interactions is one of the most promising avenues for increasing explained variability in heritable disease, this work represents a first step toward an algorithmic interaction method that preserves the information in each component. This is relevant not only because of potential improvements in model quality, but also because explicit interaction terms allow a human readable interpretation of potential interaction pathways within the disease.

\end{abstract}

\section{Introduction}\label{sec:Introduction}
Many diseases are polygenic in nature. Additionally, many biological disease pathways are known to rely on more than one single nucleotide polymorphism (SNP). Methods exist to aggregate genetic influence on a disease across large parts of the genome, primarily polygenic risk scores (PRS), which are a linear combination of the dosage of individual SNPs. Despite this, influence resulting from interactions between is difficult to account for using a PRS due to the need for a weighting value for each SNP, which does not exist for interactions. 

Using Alzheimer's Disease (AD) as a model disease (specifically data from the Alzheimer's Disease Neuroimaging Initiative) we show that explicit encoding of two gene interaction terms benefits from more nuance used to include the information from both component SNPs as well as interaction information. This second-order encoding is a first step towards understanding combinations of genes that may have an outsized impact when compared to the constituent genes considered individually, potentially due to biological pathways that depend on multiple risk alleles being present. 
\section{Methods} \label{sec:methods}
\subsection{Participant Selection and Description}
The Alzheimer's Disease Neuroimaging Initiative (ADNI) is a longitudinal multicenter study to facilitate development of clinical, imaging, genetic, and biochemical markers for the early detection and tracking of AD \cite{Veitch2019-sn, Weiner2010-ek}. Individuals were recruited from over 50 sites across the United States and Canada. Further information about ADNI can be found at \href{http://adni.loni.usc.edu/}{http://adni.loni.usc.edu/}. Primary inclusion critera included the presence of genome-wide SNP genotype data and cross-sectional amyloid PET data. 

\subsubsection{Genomic Data}
Array data was acquired from a large genome wide association study (GWAS) and filtered for standard quality control metrics, described previously \cite{Ramanan2021-nd, Saykin2015-yq}. Processed genotype files for participants from the various ADNI study phases were downloaded from the LONI web data sharing platform. Overly related samples were removed as described previously \cite{Ramanan2022-cs}. Genome-wide imputation using the TOPMed Imputation Server \cite{Das2016-ai,Taliun2021-vl} and TOPMed GRch38/hg38 build reference panel was performed separately within each batch (by GWAS array) and then merged. Monomorphic variants and SNPs with low imputation quality ($r^{2} < 0.8$) were removed. This resulted in 16,502,548 variants (8,054769 with MAF $\geq 1\%$) for 1661 individuals within the ADNI dataset. 

\subsubsection{Neuroimaging Data}
Amyloid PET imaging was performed with $^{18}$F-florbetapir (AV-45) using acquisition and processing protocols as described at \href{http://www.adni-info.org}{http://www.adni-info.org}, and with summary measures of global cortical amyloid load downloaded from the ADNI database \cite{Jagust2010-nw}. Specifically, the centiloid (CL) scale \cite{Klunk2015-pi} metric was used as the outcome of interest for global amyloid PET burden.

\subsection{Dataset Feature Selection}
To select a minimum viable feature set for proof of concept in this work, we focused first on the top 20 independent SNPs defined by largest effect size for association with clinical AD according to a recent large case/control genome wide association study \cite{BellenguezGWAS}. \textit{APOEe4} and \textit{APOEe2} were also included due to well-known associations with AD \cite{Corder1993-ub,Corder1994-vs,Ramanan2014-eg}. From this starting point, SNPs with minor allele frequency < 5\% were excluded to remove sparse features. Age, sex, and the first five genetic principal component eigenvecters were included in all models as covariates. A dataset was created from each of the following encoding methods, using interactions with \textit{APOEe4}. 

\subsection{Encoding Methods for Explicit SNP-SNP Interaction Terms}

The simplest possible encoding scheme for whether two SNPs are interacting is to assign an and operator between the SNP dosages. Because this results in either a one (if both SNPs are present) or a zero (if at least one SNP is missing), we call this binary encoding. Explicitly, this is:
\begin{equation}
    A \times B = 
    \begin{cases}
        0 & \text{if } A = 0, B = 0 \\
        1 & \text{if } A \neq 0 \text{ and } B \neq 0
    \end{cases}
    .
\end{equation} This is perhaps the most common method of consideration when attempting to define SNP-SNP interactions. However, it also fails to distinguish, which SNP is missing. Even in cases where both SNPs are present, there may be subtle differences between having one copy of each allele and two that this method fails to capture.

The next option that one may approach would likely be a linear encoding method as is here:
\begin{equation}
    A \times B =
    \begin{cases}
        A - B & \text{if } A \neq B\\
        A + B & \text{if } A = B
    \end{cases}
    .
\end{equation} This improves the variability preserved, allowing for distinction between $(A,B) = (2,2)$ and $(A,B) = (1,1)$ for example. However, this introduces a new problem, specifically that we now have $(1,0)=(2,1)$, a notable defect.

\begin{figure}
    \centering
    \begin{subfigure}[]{0.3\textwidth}
        \includegraphics[width=\textwidth]{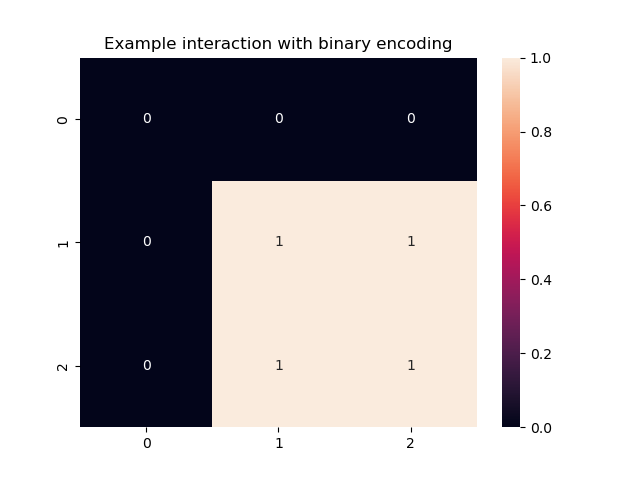}
        \caption{}
    \end{subfigure}
    \begin{subfigure}[]{0.3\textwidth}
        \includegraphics[width=\textwidth]{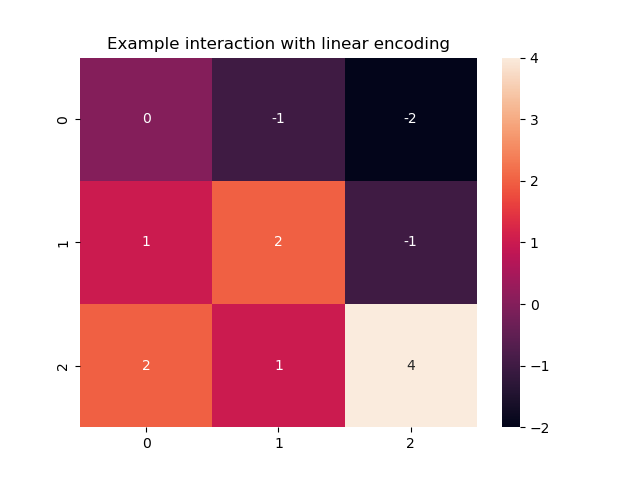} 
        \caption{}
    \end{subfigure}
    \begin{subfigure}[]{0.3\textwidth}
        \includegraphics[width=\textwidth]{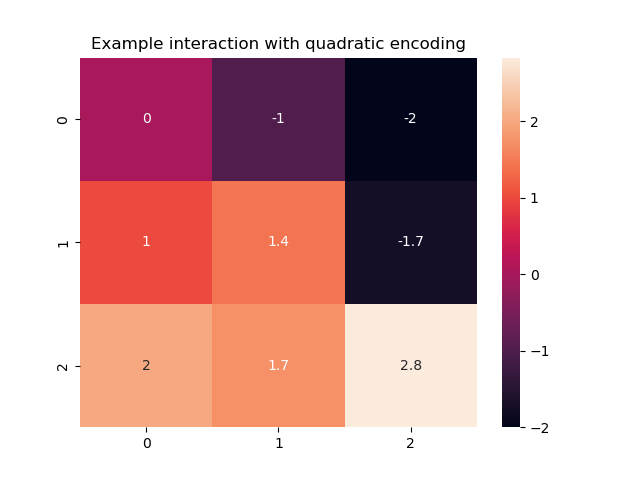}
        \caption{}
    \end{subfigure}
    \caption{Heatmaps of encoded values for all values of SNP A and SNP B. Note that as encoding grows more complex, more variance within the pairing of SNPs is observed.}
    \label{fig:heatmaps}
\end{figure}

Having discovered issues with both potential simpler methods, the next step was to add and subtract in quadrature. This allowed a preservation of all variance allowed by dosage values of zero, one or two by using the following schema: 
\begin{equation}
    A \times B = 
    \begin{cases}
        \sqrt{A^2 - B^2} & \text{if } A > B \\
        \sqrt{A^2 + B^2} & \text{if } A = B \\
        - \sqrt{|A^2 - B^2|} & \text{if } A < B
        
    \end{cases}.
\end{equation}
As with the linear encoding method, we add instead of subtract values on the diagonal to prevent it from simply going to zero. By using this method, we condense two individual SNPs into a single term that includes both the information in each individual feature and the interaction information between the two SNPs. Heatmaps illustrating differences between can be seen in Figure \ref{fig:heatmaps}.

\subsection{Machine Learning Pipeline}
A series of machine learning (ML) models were applied to each dataset, and the best performing in each was selected for comparison to the other datasets. Models were chosen for robustness to the multicolinearity introduced by our encoding method (all interactions with \textit{APOEe4} should have some collinearity, for example). Several linear regressions with varying penalization methods were used as a baseline, before moving to more advanced methods. Tree based ensemble methods were applied as the more robust models, including XGBoost and a Random Forest. Finally, a stacked meta-regression was constructed from the three best models in the sequence. All models excepting XGBoost were sourced from the scikit-learn library \cite{scikit-learn}. XGBoost was sourced from the xgboost package \cite{Chen:2016:XST:2939672.2939785}. This collection of models was run a hundred times on each dataset, and the $r^2$ correlation between actual and predicted amyloid centiloid values was collected as the primary outcome for each run. The hundred runs were then used to construct a sample distribution of $r^2$ values, boxplots of which are shown in Figure \ref{fig:boxplots}.
\section{Results}\label{sec:Results}

\subsection{Linear and Quadratic Encoding Improve Results Over Binary Encoding}
\begin{figure}[ht]
    \includegraphics[width=0.9\linewidth,center]{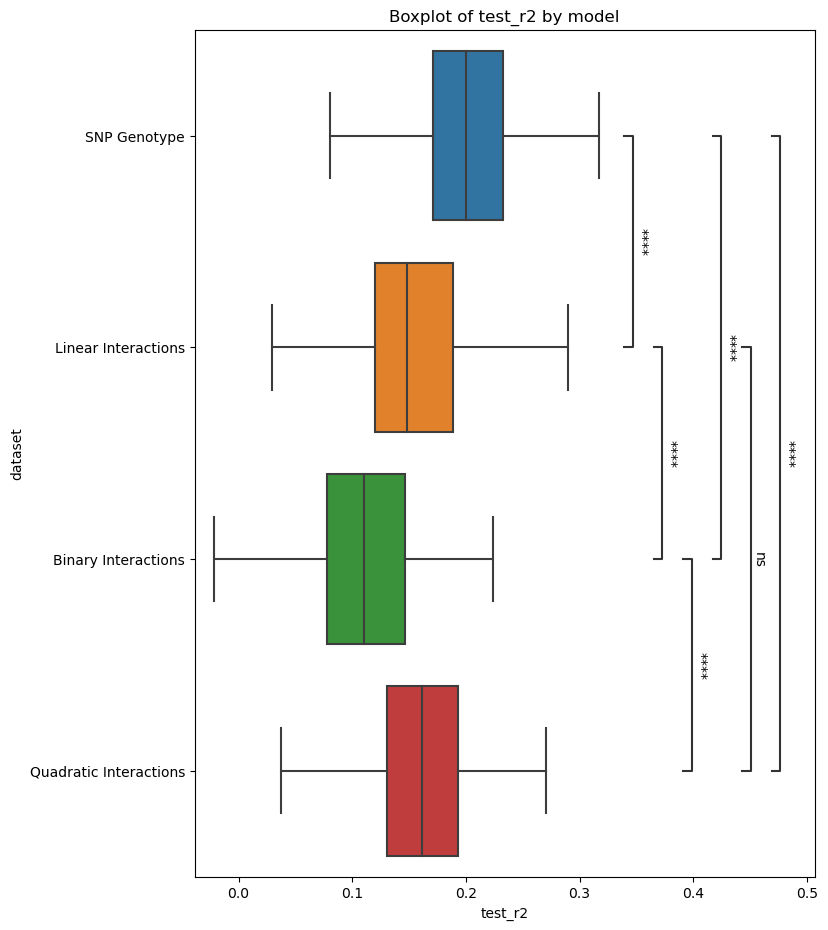}
    \caption{Boxplot comparison of best model for each interaction encoding method applied to \textit{APOEe4} interactions, as well as the raw SNP genotype.}
    \label{fig:boxplots}
\end{figure}

Both linear and quadratic encoding schemes significantly (p<0.0001) improve the correlation between predicted and actual amyloid CL values(Fig \ref{fig:boxplots}). Though the difference between linear and quadratric encoding was not significant, we still see that quadratic encoding does slightly better, likely as a result of preserving the entire variance in the interacting dosages.

\section{Discussion}\label{sec:Discussion}
We found that across various explicit encoding methods for two-way SNP interactions, those that take into account relative dosage performed significantly better than those that do not. Any interaction we applied did worse than the raw genotype dataset, though because we applied only interactions with \textit{APOEe4} this may be due to interactions that are not accounted for among the rest of the SNPs in our set. This study represents a potential first step towards explicitly encoding gene-gene interactions such that feature importance can provide biological insights in an intuitive manner.

\newpage
\printbibliography

\end{document}